\documentclass[pra,aps]{revtex4}
\usepackage{amssymb}
\usepackage{amsfonts}
\usepackage{amsmath}
\usepackage{graphicx}

\newcommand{\beq}{\begin{equation}}
\newcommand{\eeq}{\end{equation}}

\newcommand{\bea}{\begin{eqnarray}}
\newcommand{\eea}{\end{eqnarray}}

\begin{document}

\title{Quantum tomography via Non-orthogonal basis and weak values}
\author{J.J. D\'iaz }
\affiliation{Dept. de F\'{\i}sica, Universidad de Guadalajara, Revoluci\'{o}n 1500, 44420
Guadalajara, Jal., M\'exico.}
\author{I. Sainz }
\affiliation{Dept. de F\'{\i}sica, Universidad de Guadalajara, Revoluci\'{o}n 1500, 44420
Guadalajara, Jal., M\'exico.}
\author{A. B. Klimov }
\affiliation{Dept. de F\'{\i}sica, Universidad de Guadalajara, Revoluci\'{o}n 1500, 44420
Guadalajara, Jal., M\'exico.}
\affiliation{Center for Optics and Photonics, Universidad de Concepci\'on
Casilla 4016, Concepci\'on, Chile}
\date{\today }

\begin{abstract}
Using a relation between a bi-orthogonal set of equiseparable bases and the
weak values of the density matrix we derive an explicit formula for its
tomographic reconstruction completely analogous to the standard mutually
unbiased bases expansion. With the simple example of a qubit is evidenced the relationship between weak values, measured probabilities  and the separation between non-orthogonal bases.
\end{abstract}

\maketitle

\section{Introduction}

A classical state is an observable that can be completely determined through
simultaneous measurements of a certain set of (conjugate) physical
variables. On the other hand, quantum states are not observables and the
conjugate variables corresponding to non-commuting operators cannot be
measured simultaneously. The complete description of a quantum system
implies a reconstruction of a given quantum state: the so-called Quantum
State Tomography (QST) \cite{tomo} procedure. The general idea for QST of a
finite-dimensional system, of dimension $p$, is a simultaneous measurement
of $p-1$ probabilities in each of $p+1$ linearly independent bases in order
to obtain the $p^{2}-1$ real parameters describing the density matrix of the
system. It has been proved \cite{optimal} that (orthogonal) Mutually
Unbiased Bases (MUB) constitute the optimal measurement set in the case of
\emph{a priori} unknown quantum state, since the informational redundancy
among different measurements is removed. MUB tomography schemes were
successfully employed in recent experiments \cite{experiment1,experiment2}.

The situation becomes more involved if the measurement space (i.e. set of
the bases where a quantum state can be projected out) is restricted \cite%
{white}. In this case, a set of non-orthogonal bases can be used for the
density matrix reconstruction. A scheme, which preserves most of the standard
MUB tomography features \cite{pra}, is based on equidistant non-orthogonal
bases \cite{petal}, and characterized by a separation parameter $\lambda $ ($%
\lambda =0$ corresponds to the orthogonal case). In the frame of this
approach the density matrix is expanded in terms of the projectors over
specific (non-orthogonal) bases and the expansion coefficients are the
probabilities measured in the corresponding bi-orthogonal bases. An
alternative reconstruction method involving non-orthogonal measurements
(within the positive operator valued measure approach) was discussed in \cite%
{aldo}.

Unfortunately, the above mentioned non-orthogonal tomographic schemes have a
drawback: the probabilities associated to every independent element are
scaled by a factor  $\sim (1-\lambda )^{-1}$, so that when the basis
elements are close it becomes singular. This singularity is apparent in the
sense that  it disappears if exact probalitities are substituted in the
reconstruction equation. However, since (exprimentally) the  tomography is
performed with estimated probabilities, the statistical errors lead to large
deviations from the real density matrix as $\lambda \rightarrow 1$, even in
the case of perfect measurements.

Algebraically, non-orthogonal basis vectors are eigenstates of some
non-Hermitian operators. Thus, tomography in non-orthogonal bases can be
associated with measurements of non-Hermitian operators \cite%
{lundeen2,poghosyan}, which naturally appear in reconstruction schemes via
weak measurements \cite{lundeen2,johansen,salvail,elefante,lundeen1} (for
experimental implementation see Ref. \cite{lundeen2,salvail}). So we find that,
non-orthogonal tomography should be connected to the concept of weak
measurements and the two-vector formalism (assignment of
pre-and-post-selection ensembles) \cite{twostate,reznik}, that combined lead
to the notion of the weak value of an observable \cite{weak} (see \cite%
{kofman} and references therein). Numerous applications of weak values in
foundations of quantum mechanics \cite%
{weak,twostate,found,lundeen1,hofmann2010,hofmann2012}, superluminal light
propagations \cite{super} and quantum metrology \cite{metrology} have
attracted considerable attention in recent years. On the other hand, since
weak measurements do not \textquotedblleft completely\textquotedblright\
disturb the quantum state (in the sense of wave function collapse), they
seem to be naturally related to projections onto mutually non-orthogonal
states.

Here, we propose a quantum tomography scheme for finite-dimensional systems
of prime dimensions with equi-separable (non-othogonal) bases \cite{petal},
\cite{pra} using the advantage of employing a complete set of bi-orthogonal
MUBs. We present an \textit{explicit} reconstruction expression for the
density matrix, which is an analogy to the standard (orthogonal) MUB
expansion, except that it is spanned by the bi-orthogonal projectors and the
role of the measured probabilities is played by the weak values of the
density matrix. On a single qubit example we show how the weak values and
the measured probabilities are related in terms of rotations in the Bloch
sphere.

\section{Equidistant non-orthogonal bases}

In this Section we briefly summarize previous results obtained on the non-orthogonal
bases \cite{petal}, \cite{wong} and bi-orthogonal MUBs \cite{pra,jpm}. Given
a quantum system of dimension $p$, where $p\geq 3$ is a prime number ($p=2$
is discussed below), a set of $p$ non-orthogonal equidistant bases $%
\{\left\vert \psi _{m}^{s}\right\rangle ,m=0,...,p-1\}$, where the
super-index $s=1,\ldots ,p$ labels the bases, can be constructed. The
equidistant condition reads as
\begin{equation}
|\langle \psi _{m}^{s}|\psi _{n}^{s}\rangle |=(1-\lambda )\delta _{{mn}%
}+\lambda ,
\end{equation}%
where $-(p-1)^{-1}\leq \lambda \leq 1$ is the separation parameter between
elements of the same basis. It is remarkable that a set of $p$
bi-orthogonally unbiased equidistant bases $\{\left\vert \phi
_{m}^{s}\right\rangle ,s=1,\ldots ,p\}$, with separation $\eta =-\lambda
/[1+(p-2)\lambda ]$, can be found, where the unbiasedness condition reads as
\begin{equation}
|\langle \phi _{m}^{t}|\psi _{n}^{s}\rangle |^{2}=\frac{\delta _{st}\delta
_{nm}}{\mu }+\frac{1-\delta _{st}}{\mu p},  \label{unbiased}
\end{equation}%
with
\begin{equation}
\mu =\frac{1+(p-2)\lambda }{(1-\lambda )(1+(p-1)\lambda )}.
\end{equation}%
In particular, one has $\langle \phi _{m}^{s}|\psi _{n}^{s}\rangle =\delta
_{nm}/\sqrt{\mu }$.

The elements of the basis $\{\left\vert \psi _{m}^{p}\right\rangle \}$ can
be considered as eigenstates of a non-unitary cyclic operator $Z$, so that $%
Z^{t}\left\vert \psi _{m}^{p}\right\rangle =\omega ^{tm}\left\vert \psi
_{m}^{p}\right\rangle $, where $\omega =e^{2i\pi /p}$, $Z^{p}=\text{ I,}$ hence
\begin{equation}
Z^{t}=\sum_{k=0}^{p-1}\omega ^{tk}\tilde{P}_{k}^{p},  \label{z}
\end{equation}%
where we have introduced the bi-orthogonal normalized projectors $\tilde{P}%
_{k}^{s}=\sqrt{\mu }\left\vert \psi _{k}^{s}\right\rangle \left\langle \phi
_{k}^{s}\right\vert,~s=1,\ldots,p $ satisfying the following decomposition
of identity \cite{jpm,wong},
\begin{equation*}
\text{ I }=\sum_{k=0}^{p-1}\tilde{P}_{k}^{s},
\end{equation*}%
and all the operations are done $\mod p.$

The elements of the $s$-th basis, for $s\neq p$, are eigenstates of the set
of non-unitary cyclic operators which spectral bi-orthogonal decomposition
is given by
\begin{equation}
Z^{t}X^{r}=\sum_{k=0}^{p-1}\omega ^{(2^{-1}t-k)r}\tilde{P}_{k}^{s},
\label{y}
\end{equation}%
where $s=tr^{-1}$, $r,t=1,\ldots p-1,$ and the unitary, cyclic, shift
operators are defined by
\begin{equation}
X^{r}=\sqrt{\mu }\sum_{k=0}^{p-1}\left\vert \psi _{k+r}^{p}\right\rangle
\left\langle \phi _{k}^{p}\right\vert .  \label{x}
\end{equation}%
The bi-orthogonal bases are eigenstates of the correspondent adjoint
operators \cite{wong}.

The last, $p+1$-th basis required for the tomographic expansion is
orthonormal and its elements are the eigenstates of the unitary operators $%
X^{r}$. From now on, we will label this basis with the upper index $0$:
\begin{equation}
X^{r}\left\vert \psi _{m}^{0}\right\rangle =\omega ^{-rm}\left\vert \psi
_{m}^{0}\right\rangle .  \label{x basis}
\end{equation}

Explicit relations between all these bases can be consulted in Ref. \cite%
{jpm}. In particular, the overlap between the element $\left\vert \psi
_{0}^{0}\right\rangle $ and the bases $\{\left\vert \psi
_{m}^{s}\right\rangle \},\{\left\vert \phi _{m}^{s}\right\rangle \}$ has a
constant absolute value,
\begin{eqnarray}
|\langle \psi _{0}^{0}|\psi _{m}^{s}\rangle |^{2} &=&\frac{1+(p-1)\lambda }{p%
},  \notag \\
|\langle \psi _{0}^{0}|\phi _{m}^{s}\rangle |^{2} &=&\frac{1+(p-1)\eta }{p},
\label{base01}
\end{eqnarray}%
while for the rest of elements $\left\vert \psi _{n}^{0}\right\rangle ,n\neq
0$, the overlap is given by
\begin{equation}
|\langle \psi _{n}^{0}|\psi _{m}^{s}\rangle |^{2}=\frac{1-\lambda }{p},\quad
|\langle \psi _{n}^{0}|\phi _{m}^{s}\rangle |^{2}=\frac{1-\eta }{p}.
\label{base02}
\end{equation}%
Therefore, $\left\vert \psi _{0}^{0}\right\rangle $ has a very specific property:
all the components of the equidistant bases $\{\left\vert \psi
_{k}^{s}\right\rangle ,s=1,\ldots ,p\}$ approach to $\left\vert \psi
_{0}^{0}\right\rangle $ in the near parallel limit, $\lambda \rightarrow 1;$
while the bi-orthogonal bases are concentrated in the hyperplane orthogonal
to this state.

\section{Weak values and quantum tomography}

The monomials $\{Z^{t}X^{r}\}$, for $t,r=0,\ldots ,p-1$ (where $t=r=0$
corresponds to the identity operator), form a complete operational basis that
allows to reconstruct an arbitrary operator acting in the $p$ dimensional
Hilbert space. In particular, the density matrix $\hat{\rho}$ can be
expanded as follows
\begin{equation}
\hat{\rho}=\sum_{t,r=0}^{p-1}c_{tr}Z^{t}X^{r},  \label{densidad}
\end{equation}%
were the coefficients $c_{tr}$ are given in terms of the expectation values
of the non-Hermitian operators $\{X^{r\dagger }Z^{-t}\}$:
\begin{equation}
c_{tr}=\dfrac{1}{p}\text{Tr}[\hat{\rho}X^{r\dagger }Z^{-t}]=\frac{\langle
X^{r\dagger }Z^{-t}\rangle }{p},  \label{ctr}
\end{equation}%
The sets of operators $\{Z^{t}X^{r}\}$, and $\{X^{r\dagger }Z^{-t}\}$\ are
reciprocal \cite{hofmann2010},
\begin{eqnarray}
\mathit{Tr}\{X^{r}X^{r^{\prime }\dagger }\} &=&p\delta _{r,r^{\prime }},
\notag \\
\mathit{Tr}\{Z^{t}Z^{-t^{\prime }}\} &=&p\delta _{t,t^{\prime }},  \notag
\end{eqnarray}%
which grants the reconstruction Eq. (\ref{densidad}).

The expansion coefficients $c_{tr}$ have a peculiar interpretation. First of
all, $c_{00}=1/p$ due to the normalization condition; the coefficient of the
unitary terms, $X^{r}$ are
\begin{equation*}
c_{0r}=\frac{1}{p}\sum_{k=0}^{p-1}\omega ^{-kr}p_{0k},
\end{equation*}%
where $p_{0k}=\left\langle \psi _{k}^{0}\right\vert \hat{\rho}\left\vert
\psi _{k}^{0}\right\rangle $ is the probability of detecting elements of the
orthonormal basis $\{\left\vert \psi _{k}^{0}\right\rangle \}$;

The coefficients of the non-unitary elements in the expansion Eq. (\ref%
{densidad}) have the following form,
\begin{eqnarray}
c_{t0} &=&\frac{1}{p}\sum_{k=0}^{p-1}\omega ^{-tk}W_{k}^{p}(\rho ),
\label{c0} \\
c_{tr} &=&\frac{\omega ^{2^{-1}tr}}{p}\sum_{k=0}^{p-1}\omega
^{rk}W_{k}^{tr^{-1}}(\rho ),\quad r\neq 0,  \label{ct}
\end{eqnarray}%
where
\begin{equation}
W_{k}^{s}(\rho )=\frac{\left\langle \phi _{k}^{s}\right\vert \hat{\rho}%
\left\vert \psi _{k}^{s}\right\rangle }{\langle \phi _{k}^{s}|\psi
_{k}^{s}\rangle },\quad s=1,\ldots ,p,  \label{weak}
\end{equation}%
is the weak value of the density matrix for the initial state $\left\vert
\psi _{k}^{s}\right\rangle $ and post-selected state $\left\vert \phi
_{k}^{s}\right\rangle $ (in what follows we will write $W_{k}^{s}$ for
short). The concept of the weak value of the density matrix Eq. (\ref{weak})
in a non-normalized form (usually called the Dirac or Kirkwood distribution)
has been used in the context of quantum state tomography via weak
measurements \cite%
{lundeen1,hofmann2010,lundeen2,johansen,salvail,elefante,reznik}. It was
also argued in \cite{hofmann2010,hofmann2012} that such a weak value is a
complex number corresponding to the joint probability of measurements that
cannot be done simultaneously, i.e. projection into $\left\vert \psi
_{k}^{s}\right\rangle \left\langle \psi _{k}^{s}\right\vert $ and $%
\left\vert \phi _{k}^{s}\right\rangle \left\langle \phi _{k}^{s}\right\vert $%
.

Substituting the spectral decomposition of monomials $Z^{t}X^{r}$ into Eq. (%
\ref{densidad}) we arrive to the following reconstruction equation
\begin{equation}
\hat{\rho}=\sum_{k=0}^{p-1}p_{0k}P_{k}^{0}+\sum_{s=1}^{p}%
\sum_{k=0}^{p-1}W_{k}^{s}\tilde{P}_{k}^{s}-\text{ I },  \label{tomo}
\end{equation}%
where $P_{k}^{0}=\left\vert \psi _{k}^{0}\right\rangle \left\langle \psi
_{k}^{0}\right\vert $ are the projectors onto the orthogonal basis Eq.(\ref%
{x basis}). Eq.(\ref{tomo}) is our main result an it exhibits how the weak
values of the density matrix naturally appear in the frame of the
bi-orthogonal MUB tomography. It should be stressed that the above reconstruction equation does not contain any divergent factors (in the limit $\lambda\rightarrow1$) proper for non orthogonal projective tomography \cite{pra,aldo}. In the orthogonal limit, $\lambda =0$, the
weak values become true probabilities, and the $\tilde{P}_{k}^{s}$ are
transformed into orthogonal projectors so that the standard MUB reconstruction
expression \cite{optimal} is recovered. It can be easily shown that if the
density matrix expansion Eq.(\ref{tomo}) is performed on $\tilde{P}%
_{k}^{s\dagger }$ instead of $\tilde{P}_{k}^{s}$, then the corresponding
coefficients are $W_{k}^{s\ast }$, i.e.%
\begin{equation}
\hat{\rho}=\sum_{k=0}^{p-1}p_{0k}P_{k}^{0}+\sum_{s=1}^{p}%
\sum_{k=0}^{p-1}W_{k}^{s\ast }\tilde{P}_{k}^{s\dagger }-\text{ I .}
\label{tomo1}
\end{equation}%
The above equations ensure the hermiticity of the expansions Eqs.(\ref{tomo}%
)-(\ref{tomo1}).

It is worth noting that the weak values Eq.(\ref{weak}) satisfy the
normalization condition
\begin{equation*}
\sum_{k=0}^{p-1}W_{k}^{s}=1,
\end{equation*}%
in a complete analogy with the sum of probabilities $%
\sum_{k=0}^{p-1}p_{0k}=1$ measured in the orthogonal basis $\{\left\vert
\psi _{k}^{0}\right\rangle \}$.

Formally, the reconstruction equation (\ref{tomo}) depends on $(2p+1)(p-1)$
real parameters since the weak values $W_{k}^{s}$ are complex numbers.
However, such redundancy is only apparent due to specific relations between
the weak values and their complex conjugates. These relations are
straightforward to obtain by equalling matrix elements of Eq.(\ref{tomo})
and Eq.(\ref{tomo1}) in the orthogonal basis $\{\left\vert \psi
_{k}^{0}\right\rangle \}$:

\begin{widetext}
\bea
\sum_{k=0}^{p-1}\omega^{-(n-m)k}\left[2i\left(1-\lambda\delta_{m0}\right)\textrm{Im}W^p_k- p\lambda\delta_{m0}W^{p*}_k\right]=\nonumber\\
\sum_{s=1}^{p-1}\sum_{k=0}^{p-1}\omega^{(2s)^{-1}(n-m)[2k-(n+m)]}\left[2i\left(1-\lambda\delta_{m0}\right)\textrm{Im}W^s_k - p\lambda\delta_{m0}W^{s*}_k\right],\nonumber
\eea
\end{widetext}where $m=0,\ldots ,p-2$, $n=1,\ldots ,p-1$ and $n>m$. There
are $p(p-1)/2$ complex conditions, so that the total number of real
parameters required for the reconstruction in Eq.(\ref{tomo}) is reduced
from $(2p+1)(p-1)$ to the $p^{2}-1$ as it should be.

\section{Dimension two}

The general expansion Eq.(\ref{y}) is not valid in the special case of
dimension two (the quantity $2^{-1}$ is undefined). In this Section we
present explicit reconstruction equations for $p=2$ and discuss the
posiibility of their experimental implementation. Let us consider a basis
constituted by two states $\{\left\vert \psi _{0}^{2}\right\rangle
,\left\vert \psi _{1}^{2}\right\rangle \}$ with the overlap condition $%
\langle \psi _{0}^{2}|\psi _{1}^{2}\rangle =\lambda $, the corresponding
(normalized) bi-orthogonal basis is defined by
\begin{eqnarray}
\left\vert \phi _{0}^{2}\right\rangle &=&\frac{1}{\sqrt{1-\lambda ^{2}}}%
\left( \left\vert \psi _{0}^{2}\right\rangle -\lambda \left\vert \psi
_{1}^{2}\right\rangle \right) ,  \label{2b} \\
\left\vert \phi _{1}^{2}\right\rangle &=&\frac{1}{\sqrt{1-\lambda ^{2}}}%
\left( \left\vert \psi _{1}^{2}\right\rangle -\lambda \left\vert \psi
_{0}^{2}\right\rangle \right) ,  \notag
\end{eqnarray}%
where $\mu =(1-\lambda ^{2})^{-1}$, and $\langle \phi _{0}^{2}|\phi
_{1}^{2}\rangle =-\lambda $. The cyclic non-unitary operator $Z$ is then
given by the bi-orthogonal spectral decomposition
\begin{equation}
Z=\frac{1}{\sqrt{1-\lambda ^{2}}}\left( \left\vert \psi
_{0}^{2}\right\rangle \left\langle \phi _{0}^{2}\right\vert -\left\vert \psi
_{1}^{2}\right\rangle \left\langle \phi _{1}^{2}\right\vert \right) .  \notag
\end{equation}%
The unitary shift operator $X$ has the form
\begin{equation}
X=\frac{1}{\sqrt{1-\lambda ^{2}}}\left( \left\vert \psi
_{0}^{2}\right\rangle \left\langle \phi _{1}^{2}\right\vert +\left\vert \psi
_{1}^{2}\right\rangle \left\langle \phi _{0}^{2}\right\vert \right) ,  \notag
\end{equation}%
and their eigenstates $|\psi _{0,1}^{0}\rangle =\left( \left\vert \psi
_{0}^{2}\right\rangle \pm \left\vert \psi _{1}^{2}\right\rangle \right) /%
\sqrt{2(1\pm \lambda )}$ are orthonormal.

A bi-orthogonal unbiased\ to $\{\left\vert \phi _{0,1}^{2}\right\rangle \}$
basis $|\psi _{0,1}^{1}\rangle =\left( \left\vert \psi _{0}^{2}\right\rangle
\pm i\left\vert \psi _{1}^{2}\right\rangle \right) /\sqrt{2},\langle \psi
_{0}^{1}|\psi _{1}^{1}\rangle =1$ is formed by eigenstates of the operator $%
ZX$, which spectral decomposition is
\begin{equation}
ZX=\frac{i}{\sqrt{1-\lambda ^{2}}}\left( \left\vert \psi
_{0}^{1}\right\rangle \left\langle \phi _{0}^{1}\right\vert -\left\vert \psi
_{1}^{1}\right\rangle \left\langle \phi _{1}^{1}\right\vert \right) .  \notag
\end{equation}

The operators $X,Z,ZX$ and the identity form a complete set of linearly
independent operators, so that the density matrix can be expanded as
\begin{equation}
\hat{\rho}=\frac{\text{ I }}{2}+c_{01}X+c_{10}Z+c_{11}ZX.  \label{rho2}
\end{equation}%
In this particular case $Z^{-1}=Z$, and the expansion coefficients are given
by
\begin{eqnarray}
c_{01} &=&\frac{1}{2}\text{Tr}\left( \hat{\rho}X^{\dag }\right) =\frac{1}{2}%
(p_{00}-p_{01}),  \notag \\
c_{10} &=&\frac{1}{2}\text{Tr}\left( \hat{\rho}Z\right) =\frac{1}{2}%
(W_{0}^{2}-W_{1}^{2}),  \notag \\
c_{11} &=&\frac{1}{2}\text{Tr}\left( \hat{\rho}X^{\dag }Z\right) =\frac{i}{2}%
(W_{1}^{1}-W_{0}^{1}),  \notag
\end{eqnarray}%
where the probabilities $p_{0k}$ and the weak values $W_{k}^{s}$ are defined
as in the previous Section. Using the spectral decomposition of operators $%
X,Z,ZX$ and the completeness relations $p_{00}+p_{01}=W_{0}^{s}+W_{1}^{s}=1$
for $s=1,2$, the reconstruction equation for the density matrix is given by
Eq.(\ref{tomo}). There is a single (complex) condition imposed on $W_{0}^{2}$
and $W_{0}^{1}$,

\begin{eqnarray}
&&(1+\lambda )\left[ W_{0}^{2\ast }+iW_{0}^{1\ast }\right]  \notag \\
&=&(1-\lambda )\left[ W_{0}^{2}+iW_{0}^{1}\right] +\lambda (1+i).
\label{con2}
\end{eqnarray}

There is a simple scheme for obtaining the weak values $W_{k}^{1,2},k=0,1$
requiered for the reconstruction Eq.(\ref{rho2}).

First, let us choose a Pauli matrix as a unitary operator: $X=\sigma
_{z}=\left\vert \psi _{0}^{0}\right\rangle \left\langle \psi
_{0}^{0}\right\vert -\left\vert \psi _{1}^{0}\right\rangle \left\langle \psi
_{1}^{0}\right\vert $, so that the probabilities involved in the coefficient
$c_{01}$ are $p_{0k}=\left\langle \psi _{k}^{0}\right\vert \hat{\rho}%
\left\vert \psi _{k}^{0}\right\rangle ,$ $k=0,1$. If we define the
observables
\begin{equation}
\hat{M}_{k}^{y}(\alpha )=R_{y}((-1)^{k}\alpha )\sigma _{z}R_{y}^{\dag
}((-1)^{k}\alpha ),  \label{Mk}
\end{equation}%
for $k=0,1$, and $R_{y}(\alpha )=\exp (i\alpha \sigma _{y}/2)$, which are
rotations of $\sigma _{z}$ around the $y$ axis, such that $\lambda =\cos
\alpha /2$, then, the weak values $W_{k}^{2}$ are obtained by post-selecting
the states $\left\vert \psi _{k}^{2}\right\rangle $ after the weak
measurement of the observables $\hat{M}_{k}^{y}(\alpha )$
\begin{equation}
W_{k}^{2}=\frac{\mu }{2}-\frac{\mu }{2}\left\langle \psi _{k}^{2}\right\vert
\hat{M}_{k}^{y}(\alpha )\hat{\rho}\left\vert \psi _{k}^{2}\right\rangle .
\end{equation}%
Observe, that the states $\left\vert \psi _{k}^{2}\right\rangle $ can be
obtained by rotating $\left\vert \psi _{0}^{0}\right\rangle $ in the
direction opposite that used in Eq.(\ref{Mk}): $\left\vert \psi
_{k}^{2}\right\rangle =R_{y}^{\dag }((-1)^{k}\alpha )\left\vert \psi
_{0}^{0}\right\rangle $.

Similarly, $W_{k}^{1},k=0,1$ can be accessed by post-selecting the states $%
\left\vert \psi _{k}^{1}\right\rangle =R_{x}^{\dag }((-1)^{k}\alpha
)\left\vert \psi _{0}^{0}\right\rangle $ after the weak measurement of the
observable
\begin{equation}
\hat{M}_{k}^{x}(\alpha )=R_{x}((-1)^{k}\alpha )\sigma _{z}R_{x}^{\dag
}((-1)^{k}\alpha ),
\end{equation}%
where $R_{x}(\alpha )=\exp (i\alpha \sigma _{x}/2)$:
\begin{equation}
W_{k}^{1}=\frac{\mu }{2}-\frac{\mu }{2}\left\langle \psi _{k}^{1}\right\vert
\hat{M}_{k}^{x}(\alpha )\hat{\rho}\left\vert \psi _{k}^{1}\right\rangle .
\end{equation}%
Nevertheless, it follows from the completeness relation $%
W_{0}^{s}+W_{1}^{s}=1$, for $s=1,2$ that only the observable $\hat{M}%
_{0}^{y}(\alpha )$ (or $\hat{M}_{1}^{y}(\alpha )$) is requiered to obtain $%
W_{0}^{2},W_{1}^{2}$ (the same happens with $W_{0}^{1},W_{1}^{1}$). Besides,
due to relation Eq. (\ref{con2}) only one of the weak values, either $W_{0}^{1}$
or $W_{0}^{2}$ should be determined experimentally. Thus, in this
reconstruction protocol only two observables are required for the complete
determination of an unknown state: strongly measured $\sigma _{z}$ to obtain
$p_{0k}$, and, for instance, the observable $M_{0}^{y}$ which is weakly
measured and post-selected in a corresponding state. Let us remember, that it is
required to measure three observebales in the framework of the standard
orthogonal MUB tomography.

It is worth discussing the geometrical meaning of the weak values $W_{k}^{s}$%
, $k=0,1$, $s=1,2$. Let us express them in a single equation as follows
\begin{equation}
W_{k}^{s}=\sqrt{\mu }\left\langle \phi _{k}^{s}\right\vert \hat{\rho}%
\left\vert \psi _{k}^{s}\right\rangle =\mu \left\langle \psi
_{k}^{s}\right\vert \left( \left\vert \phi _{k}^{s}\right\rangle
\left\langle \phi _{k}^{s}\right\vert \hat{\rho}\right) \left\vert \psi
_{k}^{s}\right\rangle .  \label{weak2}
\end{equation}%
The above can be interpreted as weak measurements of the operators $%
\left\vert \phi _{k}^{s}\right\rangle \left\langle \phi _{k}^{s}\right\vert $
in the initial state $\hat{\rho}$ and final, post-selected, states $%
\left\vert \psi _{k}^{s}\right\rangle $, for $k=0,1$, $s=1,2$. The operators
to be weakly measured are obtained by rotations around the $x$ (or $y$) axis
from the projector on the state $\left\vert \psi _{1}^{0}\right\rangle $
located in the south pole of the Bloch sphere, for instance,
\begin{equation}
\left\vert \phi _{k}^{1}\right\rangle \left\langle \phi _{k}^{1}\right\vert
=R_{x}((-1)^{k}\alpha )\left\vert \psi _{1}^{0}\right\rangle \left\langle
\psi _{1}^{0}\right\vert R_{x}^{\dag }((-1)^{k}\alpha ),  \notag
\end{equation}%
while the corresponding post-selection states are obtained from the state
orthogonal to $\left\vert \psi _{1}^{0}\right\rangle $ (the north pole
state) and rotated on the same angle but in opposite direction:
\begin{equation}
\left\vert \psi _{k}^{1}\right\rangle =R_{x}^{\dag }((-1)^{k}\alpha
)\left\vert \psi _{0}^{0}\right\rangle .  \notag
\end{equation}%
The rotation angle are determined by $\cos \alpha /2=\lambda $ so that
when $\alpha =\pi $ ($\lambda =0$) the measured observables coincide with
the post-selection projectors and the weak values become the standard
projection probabilities on the axes $x$ (or $y$), putting in evidence the
connection between the measured probabilities and the weak values in the
frame of the optimal tomographic reconstruction.

In summary, we have found an explicit relation between the expectation
values of a set of specific non-Hermitian operators Eq.(\ref{ctr}) and the
week values of the density matrix (the normalized Dirac distribution). Such
relation allowed us to introduce a tomographic expansion of the density
matrix, Eq.(\ref{tomo}) in terms of such week values in a form analogous to
the standard MUB expansion \cite{optimal} but free of artificial
singularities.

\end{document}